\documentclass[12pt]{article}
\usepackage[english]{babel}
\usepackage[koi8-r]{inputenc}
\usepackage{epsfig}
\usepackage{amsmath,amsthm,amssymb}
\usepackage{indentfirst}
\usepackage{graphicx}
\usepackage{multirow}
\graphicspath{{images/}}
\usepackage{cite}
\usepackage[left=30mm, right=10mm, top=20mm, bottom=20mm]{geometry}
\usepackage{indentfirst} 
\newcommand{\be}{\begin{equation}}
\newcommand{\ee}{\end{equation}}

\date{}

\begin{document}

\title{Light scalar dark matter coupled to a trace of energy-momentum tensor}

\maketitle

\begin{center}

Aleksandr Belokon$^{1, 2}$, Anna Tokareva$^{2}$

\mbox{}$^{1}${\small Physics Department, Moscow State University, 119991, Moscow, Russia}

\mbox{}$^{2}${\small Institute for Nuclear Research of Russian Academy of Sciences, 117312 Moscow, Russia}

\date{\today}

\end{center}

\begin{abstract} 

We consider dark matter represented by the light scalar field whose coupling to the ordinary matter is extremely suppressed. We assume that this interaction can be described as the coupling of the square of the field to the energy-momentum tensor. We study the effect of this interaction on the evolution of dark matter scalar, as well as bounds on the model parameters that come from the variation of fundamental constants at the BBN.

\end{abstract}

\section*{Introduction}

An interesting possibility for dark matter explanation is the light scalar dark matter. Such scalar fields with extremely low masses of order $10^{-20}$ eV are common in string theory models \cite{Witten, AxionString, Axiverse, Axion}, as well as in another Standard model (SM) extensions. This type of dark matter interacts with normal matter only gravitationally and, thus, is practically invisible for detection in experiments and observations. However, light dark matter can have other couplings to the SM particles. This would rise chances of its detection in experiments. 

The scalar field can be coupled to the SM in many different ways. Here we restrict ourselves on the quadratic in the field $\phi$ interaction which preserves the $\phi\rightarrow -\phi$ symmetry. Namely, we assume that the scalar field describing dark matter is coupled to the trace of energy-momentum tensor. 

In this work, we consider the action for the scalar field coupled to the energy-momentum tensor,
\be 
\label{eq:S}
S = \int d^4x \sqrt{-g}\left(\frac{1}{2} g^{\mu\nu} \partial_\mu \phi \partial_\nu \phi-\left(\frac{m^2}{2}+\frac{1}{\Lambda^2} T^\mu_\mu\right)\phi^2\right). 
\ee

Here we concentrate on the mass range $m=10^{-16}\div 10^{-21}$ eV \footnote{Dark matter with low masses, $m \lesssim 10^{-21}$ eV, was recently constrained from the star cluster dynamics \cite{uldmgalaxy}, galactic rotation curves \cite{rotationcurves} and SDSS Lyman-$\alpha$ forest data \cite{forest1, forest2}.} (in this range, the scalar starts to oscillate after the Big Bang Nucleosynthesis (BBN)), although we discuss larger masses too. The value of $\Lambda$ is thought to be close to the reduced Planck mass $M_{P} = 2.43\times 10^{18}$ GeV. 

The interaction \eqref{eq:S} leads to the variations of the fundamental constants, such as masses of particles and SM couplings. This variations in principle can be measured directly in different experiments such as atomic clocks (see \cite{Martins} for a recent review). But the most stringent (indirect) constraints are set by the BBN epoch since the amplitude of the field is much larger at that time. The BBN bounds on dark matter couplings were obtained in \cite{Stadnik1}. However, these studies do not account for an important effect. In the presence of coupling \eqref{eq:S}, the evolution of the scalar field differs from the case of free massive scalar because the value of $T^\mu_\mu/\Lambda^2$ can be larger than the mass term. The goal of this paper is that we compute the evolution of the scalar field before the BBN including the interaction term and set more accurate bound on this kind of dark matter. We obtained that the amplitude of the scalar field is falling before the stage of oscillations which leads to additional constraints. The reason is that for the significant part of the parameter space it is impossible to provide the correct dark matter abundance.

The paper is organized as follows. In Section \ref{sec:abouttrace} we discuss contributions of all the SM particles that are in thermal equilibrium with the radiation bath in the early Universe to the trace of energy-momentum tensor. In Section \ref{sec:aboutansol} we obtain an analytical solution for the approximate choice of the trace of energy-momentum tensor for qualitative understanding of the field dynamics. Then, in Section \ref{sec:DM} under the assumption that this scalar field forms all dark matter, we use these results to derive a constraint. In Section \ref{sec:coupling} we discuss the bounds appeared from considering the influence of the dark matter scalar on the BBN dynamic. In Section \ref{sec:afterBBN} we suggest the model that can come over this restriction. In Section \ref{sec:Conclusions} we summarize our results.

\section{Cosmological evolution of the scalar field coupled to the energy-momentum tensor}

\subsection{Trace of the energy-momentum tensor during the radiation dominated stage}

During radiation domination, the equation of state $p=\rho/3$ implies that the energy-momentum tensor is traceless, $T^{\mu}_{\mu}=\rho-3 p=0$. However, this relation is approximate. $T^{\mu}_{\mu}$ receives contributions from different sources. At temperatures higher than $100$ GeV its value is defined by the gauge trace anomaly \cite{Morozov, hotpressure, gravbar},
\be 
\label{eq:T}
T^{\mu}_{\mu} = \beta\rho, \quad \beta \sim 10^{-3}.
\ee
At lower temperatures, the main effect is provided by the particles that become non-relativistic. In this case, in hydrodynamic limit, the trace can be written down as follows \cite{kick},
\be
T^\mu_\mu(T)=\sum_{i}\frac{g_i m_i^2}{2\pi^2}\int_{m_i}^\infty\frac{\sqrt{E^2-m_i^2}}{\exp(E/T)\pm1}dE=\sum_{i}\frac{g_i}{2\pi^2}m^2_i T^2 \int_{m_i/T}^\infty\frac{\sqrt{u^2-(m_i/T)^2}}{\exp(u)\pm1}du,
\label{eq:gentrace}
\ee
where we have introduced $u=E/T$ as the integration variable in the last line; $i$ is the index of all SM particle species that are non-relativistic at the temperature $T$; $m_i$ is the particle's mass, $g_i$ is the number of degrees of freedom for the particle species; T is the temperature of the radiation bath; and the $+$ sign in the denominator applies to fermions, while the $-$ sign applies to bosons. 

In particular, at the time of the BBN, the trace is mostly determined by the annihilation of electrons and positrons. The value of $\beta=T^{\mu}_{\mu}/\rho$ varies with the temperature from $10^{-3}$ (at high temperatures) to $10^{-2}$ during annihilation of electrons and positrons. Thus, for analytical estimation of the effect of the interaction between dark matter and SM plasma, in the next section, we use a constant value of $\beta=10^{-3}$.

\begin{figure}[h!]
\center{\includegraphics[scale=0.6]{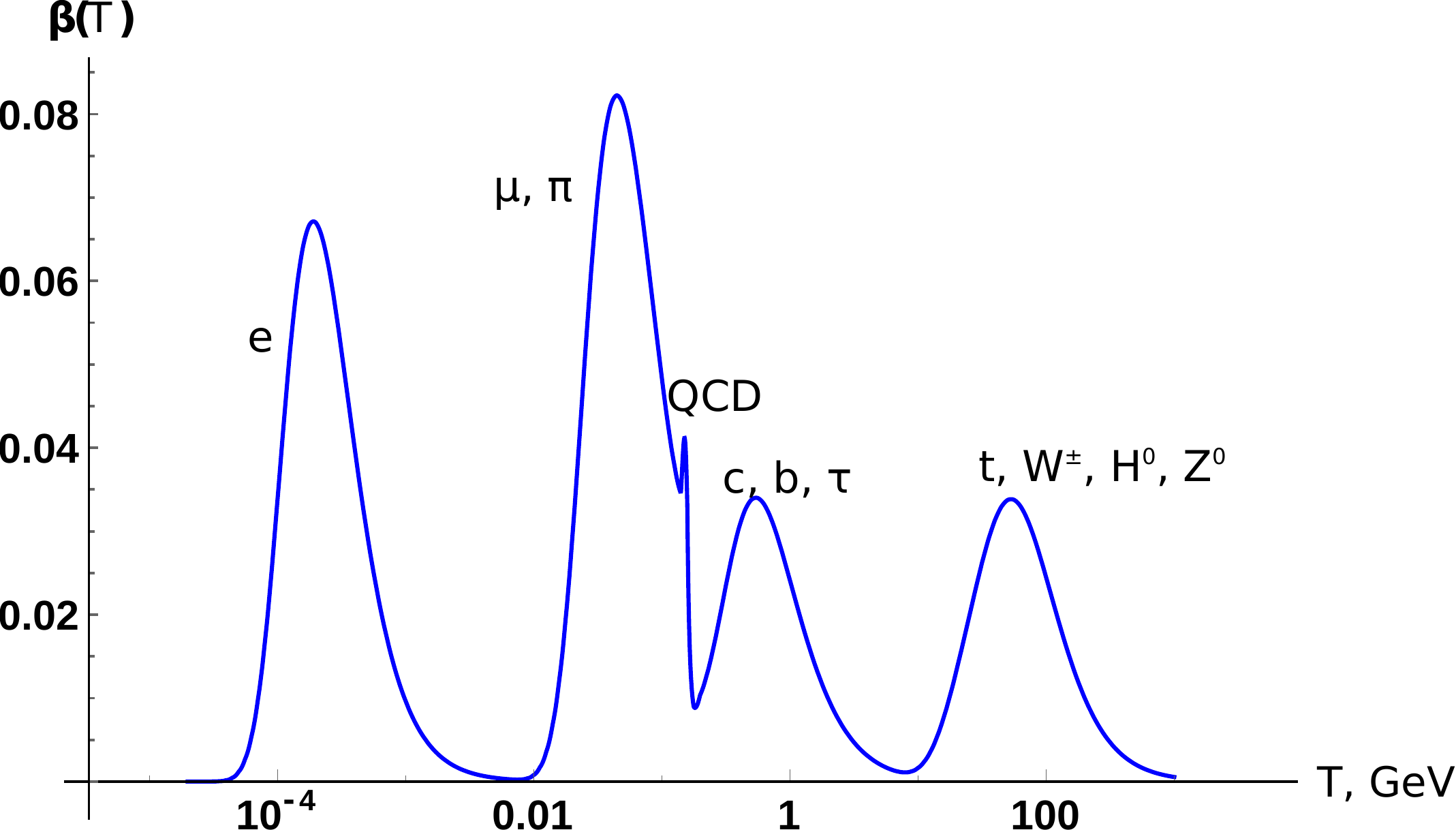}} \\
\caption{\footnotesize{The function $\beta(T)=T^\mu_\mu/\rho$ as a function of the temperature. $\beta$ deviates from zero when the temperature falls below the mass of a particle that is in thermal equilibrium with the radiation bath. It includes contributions from all the SM-particles. The discontinuity at the temperature of 170 MeV corresponds to the QCD phase transition}.}
\label{fig:beta}
\end{figure}
\label{sec:abouttrace}

\subsection{Analytic solution to the field equations}

Here we use a formula (\ref{eq:T}) for $T^{\mu}_{\mu}$, in order to study how the evolution of the scalar field is affected by the time-dependent contribution to its effective mass. As a first approximation, we take a constant value of $\beta\sim 10^{-3}$. In this case, we obtain the following equation of motion for the field $\phi$,  

\be 
\label{eq}
t^2\ddot \phi + \frac{3}{2}t\dot\phi + \left( m^2 t^2+\alpha^2 \right)\phi = 0,\quad \phi(t_0) = \Lambda,~\dot \phi(t_0) = 0,
\ee
where $\alpha=\displaystyle\frac{3\beta M^2_{Pl}}{2\Lambda^2}$. Here we set the initial conditions at some moment of time, $t_0$. If we switch off the extra coupling to the trace of energy-momentum tensor, the concrete value of $t_0<1/m$ doesn't matter since the solution is constant until $t\sim 1/m$ (here we neglect the falling solution). We show that it is not the case if the coupling to matter is included. In analogy with the axion, we interpret the initial moment as a moment of phase transition when the shift symmetry gets broken to the discrete subgroup and the scalar gains the mass and interaction terms. We assume that it happens fast, so in such a way that we can simply start the evolution from the moment $t_0$. The natural value of $\phi$ at this moment is of order $\Lambda$, since $\phi$ has a discrete symmetry.

This equation has an analytical solution in terms of Bessel functions,
 \be 
 \phi(t)=C_1 m^{3/4}t^{-1/4} J_{\frac{1}{4}\sqrt{1-16\alpha^2}}(mt)+C_2 m^{3/4}t^{-1/4} J_{-\frac{1}{4}\sqrt{1-16\alpha^2}}(mt),
 \ee
where the dimensionless constants $C_1$ and $C_2$ are defined by the initial conditions.

At small $t\ll 1/m$ the solution has a form
\be 
\phi(t) \approx \Lambda \left(\frac{t_0}{t}\right)^{\frac{1}{4}(1-\sqrt{1-16\alpha^2})},
\ee
in case of $\alpha \leq 1/4$, and
\be 
\phi(t) \approx \Lambda\left(\frac{t_0}{t}\right)^{1/4}\sin\left(\gamma\ln\left(\frac{mt}{2}+const\right)\right), \;\;\; \gamma = \frac{1}{4}\sqrt{16\alpha^2-1},
\ee
when $\alpha > 1/4$.

After $t=1/m$ the scalar in both cases starts to oscillate. The solution that corresponds to the initial conditions \eqref{eq} is,
\be
\phi\approx\Lambda(mt_0)^{\frac{1}{4}(1-\sqrt{1-16\alpha^2})}(mt)^{-3/4}\cos(mt+const),
\ee
for the case of $\alpha\leq 1/4$, and
\be
\phi\approx\Lambda(mt)^{-3/4}(mt_0)^{1/4}\cos(mt+const),
\ee
for $\alpha\geq 1/4$.

We see that the coupling of dark matter to the energy-momentum tensor drastically changes the behaviour of the field before oscillations. Namely, the solution is falling which can lead to several bounds of the model parameters. It is clear that in some cases the field value would fall so fast that it can not provide the needed amount of dark matter anymore. In the next section, we find the model parameters which still allow for the scalar field to explain all dark matter in the late Universe.

\begin{figure}[h!]
\center{\includegraphics[scale=1.94]{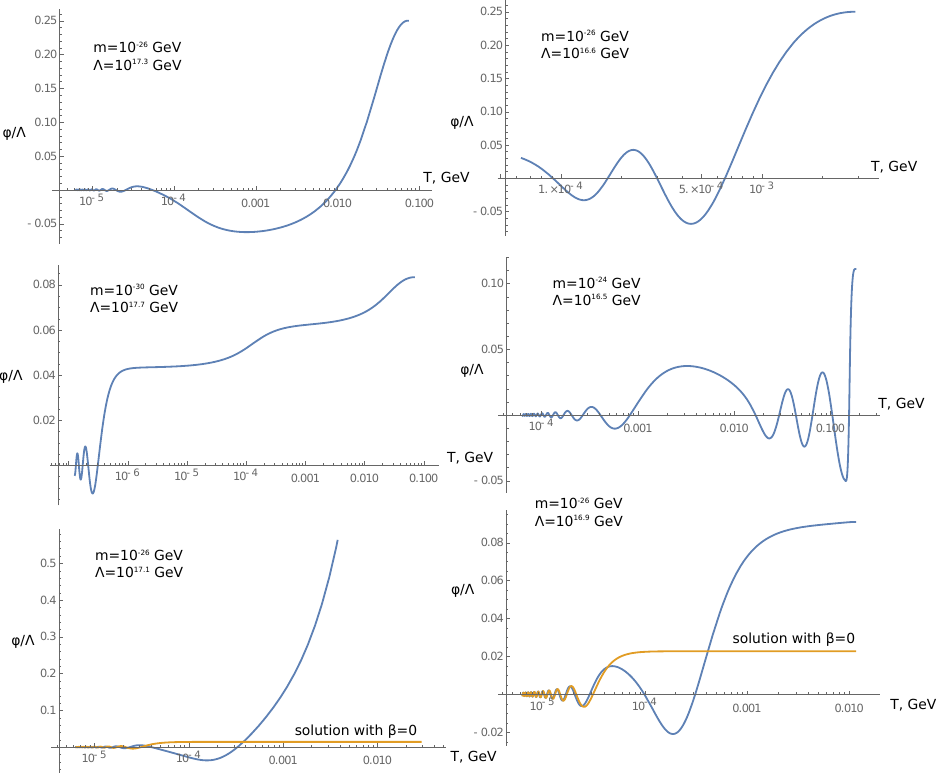}}
\caption{\footnotesize{The variety of solutions for different parameters of the model. Numerical results were obtained for the trace of the form (\ref{eq:gentrace}), see also Fig. \ref{fig:beta}. In the last two plots, we draw the solution in a model without the coupling to the matter (orange line) which produces the same amount of dark matter in the late Universe.}}
\label{fig:GraphicsSolution}
\end{figure}
\label{sec:aboutansol}

\subsection{Constraints from the dark matter production}
As we mentioned above, the extra coupling to the trace of energy-momentum tensor leads to the falling of the field's amplitude. Due to this falling, it does matter when the phase transition of the field $\phi$ has happened. As we see later, not in all cases it is possible to obtain the energy density required for this field, in order to explain dark matter.

To obtain constraints on the parameters that allow our field to be the observed dark matter, we employ the conservation of entropy $n/s=const$. Here $n$ is the number density of dark matter particles and $s$ is the total entropy density of all particles in thermal equilibrium with the radiation bath. 

At the moment $t\sim 1/m$, the field starts to oscillate and its energy density is, 
\be
\langle\rho_{\phi}\rangle\approx\frac{m^2\Lambda^2}{2}(mt_0)^{\frac{1}{2}(1-\sqrt{1-16\alpha^2})},
\ee
for case $\alpha \leq 1/4$ ($\Lambda\geq\sqrt{24\beta}M_{P}$); and
\be
\langle\rho_{\phi}\rangle\approx \frac{m^2\Lambda^2}{2}(mt_0)^{\frac{1}{2}},
\ee
when $\alpha \geq 1/4$ ($\Lambda\leq\sqrt{24\beta}M_{P}$).

Dividing these expressions by $m$, we obtain the number density $n_{\phi}$ of dark matter particles for this epoch:
\be
\label{n}
n_{\phi}\approx\frac{m\Lambda^2}{2}(mt_0)^{\frac{1}{2}(1-\sqrt{1-16\alpha^2})},
\ee
for case $\alpha \leq 1/4$ ($\Lambda\geq\sqrt{24\beta}M_{P}$); and
\be
n_{\phi}\approx \frac{m\Lambda^2}{2}(mt_0)^{\frac{1}{2}},
\ee
when $\alpha \geq 1/4$ ($\Lambda\leq\sqrt{24\beta}M_{P}$).

The entropy density $s$ at the time is 
\be
\label{s}
s=\frac{2\pi^2}{45}g_{\rho}(T_{osc})T_{osc}^3.
\ee

Here $g_{\rho}(T)$ is the number of degrees of freedom in the cosmic plasma as a function of the temperature $T$ of the radiation bath and $T_{osc}$ is the temperature at which the field starts to oscillate,
\be
T_{osc}= \sqrt{\frac{m M_{Pl}}{1.66\sqrt{g_{\rho}}}},
\label{eq:Tosc}
\ee
where $M_{Pl} = 1.2209\times 10^{19}$ GeV is the Planck mass \cite{pdg}.

The current energy density of the dark matter is characterized by density parameter $\Omega_{DM}\equiv\rho_{DM}/\rho_{crit}=0.2581$ \cite{pdg},  where $\rho_{crit}=0.53\times 10^{-5}$ $\text{GeV}/\text{cm}^3$ is the critical density. Thus, the number density $n_0$ can be evaluated as,
\be
\label{n0} 
n_0 = \rho_{crit}\Omega_{DM}/m.
\ee
The present value of the entropy density is
\be
\label{s0}
s_0=2.9\times 10^3 \; \text{cm}^{-3}.
\ee

Now let us write down the conservation of entropy for the moments $t\sim 1/m$ and today:
\be 
\frac{n_{\phi}}{s}=\frac{n_0}{s_0}.
\ee 

Substituting here (\ref{n})--(\ref{s0}) and solving this equation for $t_0$, we obtain,
\begin{equation*}
t_0(m, \Lambda) = 
 \begin{cases}
  \displaystyle \frac{1}{m} \left(\frac{4\pi^2\,\Omega_{DM}\,\rho_{crit}}{45\,s_0\,m^2\,\Lambda^2}\right)^{\frac{2}{1-\sqrt{1-\frac{24\beta M_{P}^2}{\Lambda^2}}}}, & \text{if} \quad\Lambda\geq\sqrt{24\beta}M_{P}\\
  \\
   \displaystyle \frac{1}{m}\left(\frac{4\pi^2\,\Omega_{DM}\,\rho_{crit}}{45\,s_0\,m^2\,\Lambda^2}\right)^2, & \text{if} \quad\Lambda\leq\sqrt{24\beta}M_{P}
 \end{cases}.
\end{equation*}

\begin{figure}[h!]
\center{\includegraphics[scale=1]{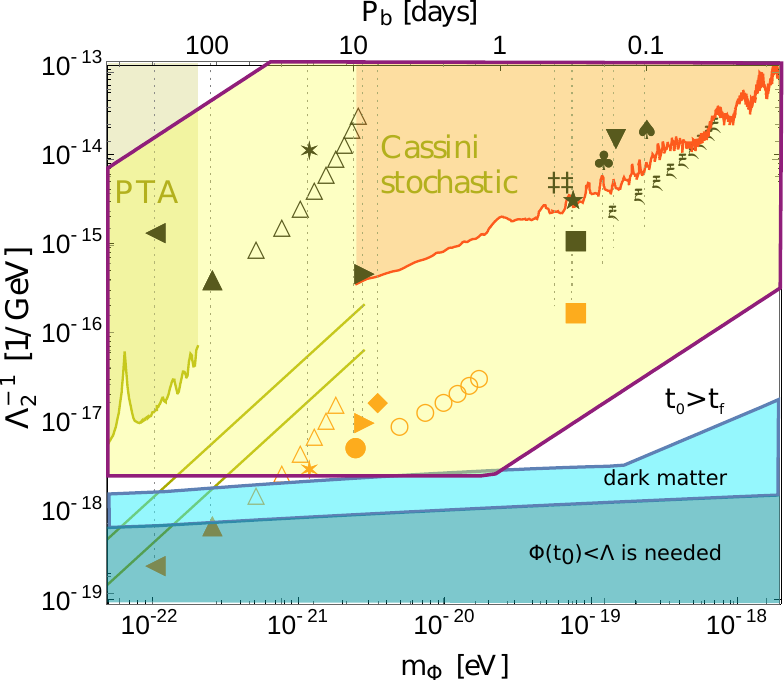}}
\caption{\footnotesize{The sensitivity of binary pulsar observations to the quadratically coupling between dark matter and ordinary matter on the example of several known systems \cite{Sibiryakov} versus constraints obtained in Section \ref{sec:DM}. Black symbols are constraints derived using the existing data on the time derivative of the orbital period of binary system $\langle\dot P\rangle$ which interacts with the dark matter background; values above the symbols are excluded. Orange symbols show the sensitivity that can be achieved if $\langle\dot P\rangle$ is measured for a given system with the accuracy $10^{-16}$. Empty symbols correspond to resonances on higher harmonics. The coloured regions of the dark matter parameter space are excluded by PTA \cite{Postnov} (olive), and Cassini bound on stochastic GW background \cite{Cassini} (red). Olive lines show future sensitivities of European Pulsar Timing Array (upper) and Square Kilometer Array (lower), as estimated in \cite{coeff}. Also, this plot shows the constraint (\ref{eq:constr}) that comes from the requirement that the field starts its evolution (i.e. gains the mass) before the BBN and neutron freeze-out (blue region); and after the BBN (yellow region), see the model constructed in Section \ref{sec:afterBBN}. In the dark blue region, in order for $\phi$ to be all dark matter, one needs $\phi(t_0)<\Lambda$}.}
\label{fig:puls}
\end{figure}

Then, using this equation and the ratio between the age of the Universe $t$ and its temperature $T$,
\be
T(t)=\sqrt{\frac{M_{Pl}}{1.66\sqrt{g_{\rho}}}}\;t^{1/2},
\label{Tt}
\ee
we can express the temperature $T_0(m, \Lambda)$ of the phase transition of the field $\phi$. Since we don't specify how the scalar $\phi$ gains its mass, we assume that the moment $T_0$ happened before the weak interaction freeze-out ($T_f\sim 3$ MeV). In this case, the details of this process do not affect the BBN. Thus,
\be
\label{eq:constr}
T_{f}\leq T_0(m, \Lambda),
\ee
which provides us with the constraints in the parameter space $(m, \Lambda)$, see Fig. \ref{fig:puls}.

\label{sec:DM}

\section{The BBN constraints on the coupling to matter}

\subsection{Variations of the fundamental constants induced by the scalar field}

The coupling of the DM field to the stress-energy tensor of ordinary matter directly leads to small variations of particle masses. For example, in the low energy SM effective action, the electron mass term contributes to $T_{\mu}^{\mu}$ as,
\be 
T_{\mu}^{\mu}(e)=m_e \bar{\psi}\psi.
\ee
Therefore, the coupling (\ref{eq:S}) leads to the shift of the electron mass,
\be 
m_e=m_e^0\left(1+\frac{\phi^2}{\Lambda^2}\right).
\ee
Note that masses of quarks are changed in a similar way. However, the variations of the proton and neutron mass and the neutron lifetime requires more involved computations because they are also affected by the QCD strong coupling scale $\Lambda_{QCD}$ \cite{Campbell}. The result computed for the SM $\beta$-function is
\be 
\Lambda_{QCD}=\Lambda_{QCD}^0\left(1-\frac{14}{27}\frac{\phi^2}{\Lambda^2}\right),
\ee
The neutron-proton mass difference, the crucial quantity for the BBN calculations, is then obtained as \cite{quarks}
\be 
m_n-m_p=(m_n-m_p)_0\left(1+w \frac{\phi^2}{\Lambda^2}\right),\quad w=f_s+\frac{2}{9}-\frac{14}{27}+1\approx 0.82.
\label{eq:mnmp}
\ee

Here $f_s=0.113 \pm 0.053$ \cite{coeff} characterises the strange quark impact on the nucleon mass difference, $2/9$ comes from the dependence of the heavy quarks ($c, b, t$) mass thresholds on the scalar field, $\phi$.

The neutron lifetime scales as $\tau_n\sim G_F^2 m_e^5$, thus
\be
\tau_n=\tau^0_n\left(1+\frac{\phi^2}{\Lambda^2}\right).
\label{eq:tn}
\ee
If these two fundamental constants, $m_n-m_p$ and $\tau_n$, differ from their SM values, the dynamics of the BBN is affected. Namely, the helium production is highly sensitive to the difference between neutron and proton mass. In the next section, we discuss the numerical results for the primordial production of $^4$He in the model with light scalar dark matter.
\label{sec:coupling}

\subsection{Numerical results for the helium production}

Calculation of the primordial helium abundance at the time of the BBN requires numerical computation of the asymptotically surviving neutron abundance $X_n\equiv\displaystyle\frac{n_n}{n_B}$, in the presence of a scalar field $\phi$. Here $n_n$, $n_B$ are neutron and baryonic number densities respectively, $n_B=n_n+n_p$. In our computations, we used an analytic description of the neutron freeze-out process given in \cite{BBN}:
\be
X_n(T)\equiv X_n(y(T))=X_n^{eq}(y)+\int\limits_0^y dy^\prime e^{y^\prime}(X_n^{eq}(y^\prime))^2 \exp(K(y)-K(y^\prime)),
\label{eq:xn}
\ee

\be
y(T)=\Delta m(\phi)/T, \quad \phi=\phi(T),
\ee
where $\Delta m(\phi)$ is the neutron-proton mass difference in the presence of the field $\phi$ which is given by (\ref{eq:mnmp}).

The equilibrium neutron abundance is 
\be
X^{eq}_n(T)=\left(1+e^{y(T)}\right)^{-1}.
\ee

The function $K(y)$ defining the rate of neutron production is given by,
\be
K(y)=-b\left(\left(\frac{4}{y^3}+\frac{3}{y^2}+\frac{1}{y}\right)+\left(\frac{4}{y^3}+\frac{1}{y^2}\right)e^{-y}\right),
\ee

\be 
b=a\left(\frac{45}{4\pi^3 g_{\rho}(T_{BBN})} \right)^{1/2}\frac{M_P}{\tau_n(\phi) \Delta m(\phi)^2}, \quad a=253,
\ee
where $T_{BBN}\simeq 6.3968\times 10^{-5}$ GeV is the moment of helium production \cite{pdg}, in terms of the temperature; and $\tau_n(\phi)$ is the neutron lifetime in presence of $\phi$ which is given by the (\ref{eq:tn}).

Thus, the mass fraction of helium in the presence of the light scalar field is given by \footnote{This approximate formula, nevertheless, provides with correct results, since the neutrons freeze-out much earlier than $t_{BBN}$ and the neutron decay does not affect the production of neutrons.}
\be
X_{^4 He}(\phi)\simeq 2 X_n(T_{BBN})\exp\left({-\int^{t_{BBN}}_{t_0(m,\,\Lambda)}\tau_n(\phi(t))^{-1}dt}\right),
\ee
where $t_{BBN}\simeq 180$ s is the time of the BBN \cite{pdg}.

The helium fraction is bounded by the Planck data \cite{Planck} as,
\be 
0.2464 \leq X_{^4 He}\leq 0.2505.
\ee

\begin{figure}[h!]
\center{\includegraphics[scale=0.7]{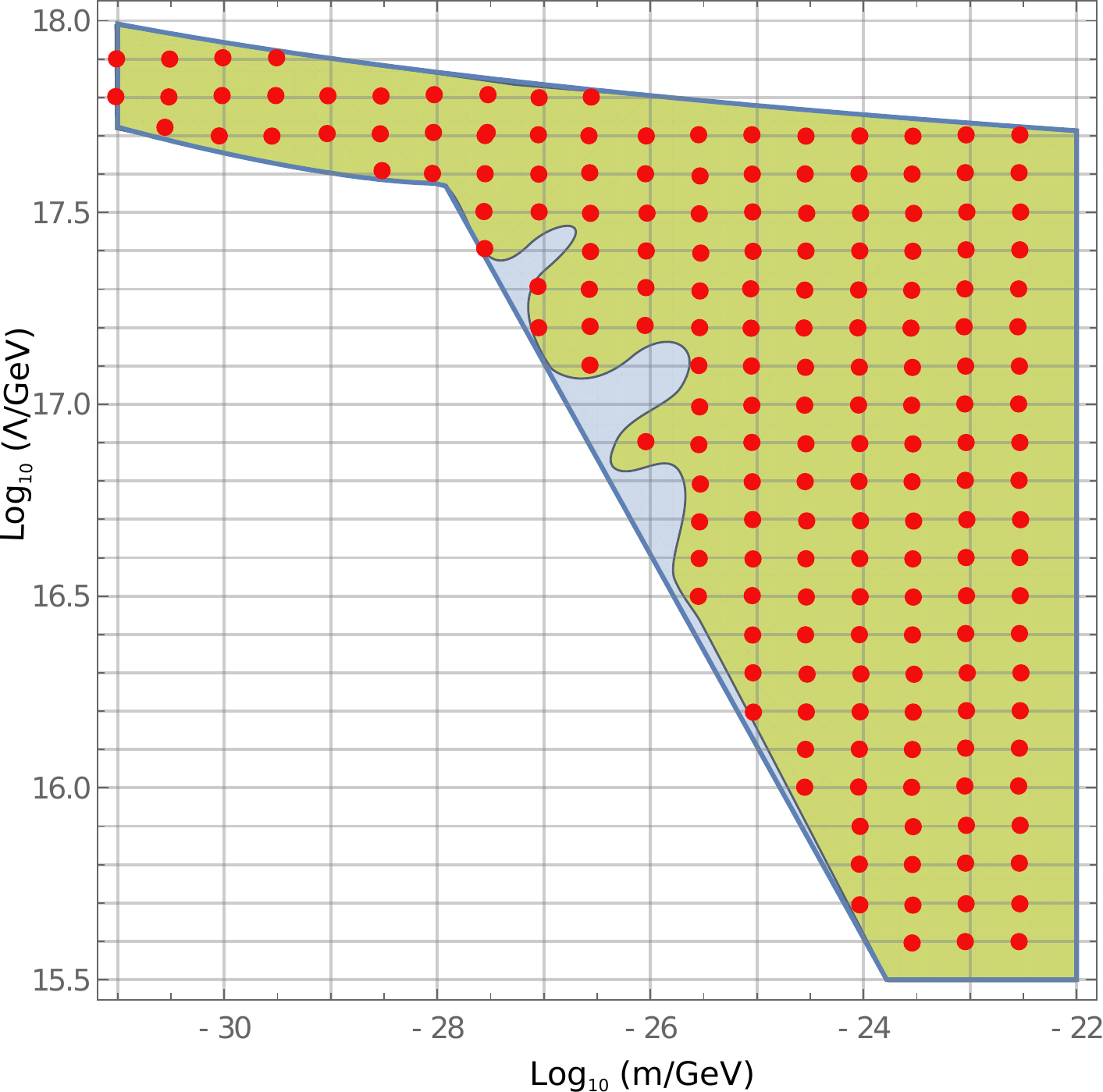}}
\caption{\footnotesize{The coloured regions of the dark matter parameter space (m, $\Lambda$) are allowed by the mentioned constraints. Blue region is a constraint derived using the existing data on the amount of dark matter in the Universe (Section \ref{sec:DM}); orange region shows the constraint from the BBN (Section \ref{sec:coupling}). Red points are the lattice nodes which were interpolated. The region upper the line on the top of the coloured region is also allowed by the BBN constraint. However, one needs $\phi<\Lambda$, in order to produce the correct amount of dark matter}}
\label{fig:constr}
\end{figure}

We computed numerically the helium production with the scalar field normalized on the dark matter abundance, for several values of $m, \Lambda$. We used a numerical solution for the field $\phi$ obtained with the realistic behaviour of the trace of energy-momentum tensor (see Fig. \ref{fig:beta}). We took the initial moment from (\ref{eq:constr}) and assume that the derivative of the field is zero. Then, we chose the initial field value in such a way that the amplitude of the late time oscillations corresponds to those of dark matter. Within this solution, we found that in most points satisfying the constraint Fig. \ref{fig:constr} the amount of helium falls into the allowed region. However, we obtained some small exclusion regions. They correspond to the masses of dark matter at which the field starts to oscillate at the time of the BBN, thus providing with non-trivial dynamics during this transition period. 

For dark matter with low masses, we obtained the bound on the scale $\Lambda$ approximately consistent with the results obtained in \cite{Stadnik2}. For dark matter which starts to oscillate during the neutron freeze-out and the BBN, our constraint is somewhat weaker than those of \cite{Stadnik2} because, in opposition to this paper, we didn't make averaging of the oscillations. The integral (\ref{eq:xn}) appears to be closer to the SM one, due to the oscillatory behaviour of the integrand, as compared to the case of the averaged field. Thus, we obtained that, in fact, the model is bounded mostly by the requirement that we set the initial conditions before the neutron freeze-out. In the next section, we construct an example of the model which allows for overcoming this constraint.

\section{Can dark matter be produced after the BBN?}
\label{sec:afterBBN}
In the previous section, we obtained that the limits coming from the condition that the scalar field forms all dark matter do not leave a window for other searches of dark matter. At first glance, the values of $m$ and $\Lambda$ available for the astrophysical probes with binary pulsars are forbidden by these constraints. Does it mean that these searches are not motivated? Here we show that it is still possible for this type of dark matter to avoid the BBN constraints, together with the constraints from the initial time of the phase transition, if dark matter is formed after the BBN. In this case, the values of fundamental constants during the BBN are the same as in the Standard model. However, in the late Universe, the constants can vary with the scalar field value providing with dark matter coupled to the mass \cite{Sibiryakov} relevant for pulsar timing searches. Here we present a concrete model which allows for avoiding the BBN constraints.

Assume the scalar field has a discrete shift symmetry and its potential has a usual form,
\be
\label{V_cos}
V(\phi)=m^2\Lambda^2\left(1-\cos\left(\frac{\phi}{\Lambda}\right)\right).
\ee
This scalar is coupled with the Standard model particles via the term
\be 
\label{int_cos}
L_{int}(\phi, T)=\kappa T_{\mu}^{\mu}(T) \left(1-\cos\left(\frac{4\phi}{\Lambda}\right)\right).
\ee
Here we take
\be 
T_{\mu}^{\mu}(T)=\beta g_{*}\frac{\pi^2}{30} T^4.
\ee

\begin{figure}
\center{\includegraphics[width=0.6\textwidth]{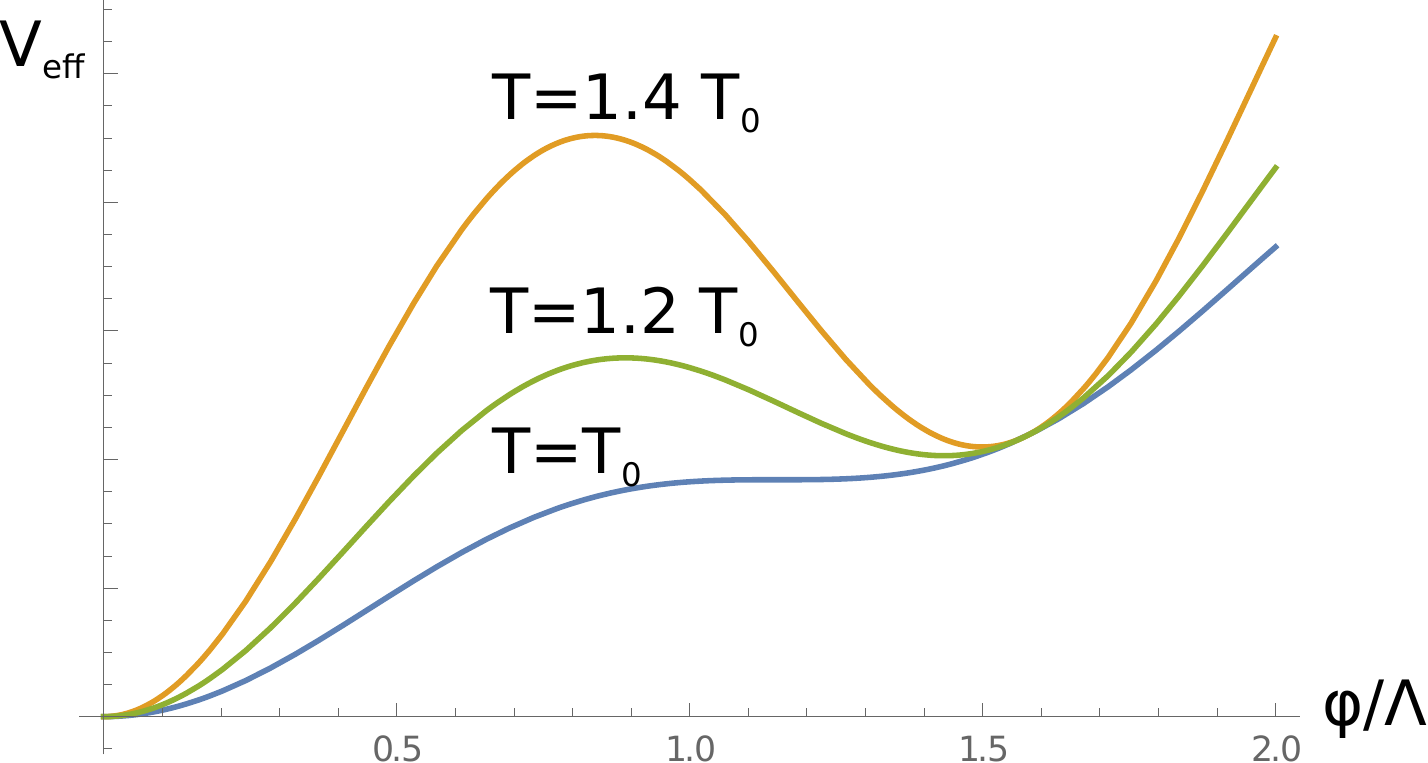}}
\caption{\footnotesize{Effective potential \eqref{V_cos}, \eqref{int_cos} for different values of temperature before the minimum disappears.}}
\label{fig:cos}
\end{figure}

In the late Universe, the minimum of the effective potential $V+L_{int}$ is $\phi=0$.  Thus, expanding the action around this minimum one obtains the model \eqref{eq} with the quadratic coupling to $T_{\mu}^{\mu}$. However, in the early Universe, it can happen that the interaction term dominates over the potential and the scalar is in another minimum, $\phi/\Lambda\simeq \pi/2$. This lasts until some critical temperature $T_0$ (see Fig. \ref{fig:cos}),
\be 
\kappa\,\beta\,g_{*}\,\frac{\pi^2}{30}\,T_0^4=0.23\,m^2\Lambda^2.
\ee
At this temperature, the minimum disappears and the scalar starts rolling down and oscillating. From the conservation of the entropy, one can derive an amount of dark matter in the present Universe. The scalar field explains all dark matter if the condition
\be 
m\,\Lambda=1.25\cdot 10^{-20}\,\kappa^{-3/2} \,\, \text{GeV}^2,
\ee
is satisfied\footnote{We assume that the field starts to oscillate immediately after the moment $T_0$. This assumpltion works if $\Lambda<4\cdot 10^{18}\sqrt{\kappa\beta}$.}. The condition that the BBN is unaffected by this scalar dark matter (we take $T_0<1$ keV, which corresponds to the time of when the BBN is finished) implies,
\be 
\kappa\beta\gtrsim 5.7\cdot 10^{-5}.
\ee
At the same time, since it is natural to have the coupling constant $\kappa<1$, we expect $\kappa\beta \lesssim 10^{-2}$. These bounds leave the allowed region for $m$ and $\Lambda$ which is shown in Fig. \ref{fig:constr}. One can see that this region is quite large and it covers values available for the observations of binary pulsars.

Let us check that, under the listed conditions, the BBN is indeed not affected by the varying fundamental constants. In this model, during the BBN the scalar is in the other minimum $\phi_{min}/\Lambda=\pi/2-0.068\,(T_0/T)^4$ where the interaction term with matter can be written as, 
\be 
L_{int}\simeq 0.01\,\kappa\,T_{\mu}^{\mu}\,\left(\frac{T_0}{T}\right)^8.
\ee
This means that the SM couplings at the temperature $T$ differ from the present ones by $\Delta g/g\sim\,0.01\alpha\,(T_0/T)^8$ which is negligible for the interesting values of $T_0\sim 1$ keV. Thus, the model \eqref{V_cos}, \eqref{int_cos} is indeed a self-consistent description of dark matter which leads to varying constants and quadratic coupling to masses in the late Universe and, at the same time, does not affect the BBN. 

\section{Conclusions}

In this work, we considered a light scalar field which is introduced to explain all dark matter in the late Universe. Analogically to the case of the axion, after the phase transition breaking its shift symmetry, this field gains a mass and a coupling to matter. We examined the case in which this coupling is universal (i.e. the field interacts with the trace of energy-momentum tensor) and quadratic in the field. We found that in the early Universe, the average value of $T_{\mu}^{\mu}$ might be much larger than the mass term in the equation of motion, providing with the drastic change of the dynamics before the oscillation period. We took into account this effect and obtained the accurate bounds on the mass and coupling to matter which allows for the field to be dark matter. We found that if the field initiated before the BBN the allowed region is, unfortunately, far from the parameters which can be probed within the observations of binary pulsars.

This kind of dark matter would affect the values of all the fundamental constants. This leads to the constraints coming from the abundance of light elements produced during the nucleosynthesis. In this work, we computed the helium production with the fundamental constants (the neutron-proton mass difference $m_n-m_p$ and the neutron lifetime $\tau_n$) varying with the field evolution. We found that, except for several small exclusion regions, the final helium abundance appears to fall into the region allowed by the Planck data. Thus, the bound connected with the very possibility to obtain the correct amount of dark matter occurred to be more restrictive. 

Unfortunately, this bound doesn't leave a possibility to probe this model in experiments and observations. However, we suggested a model that allows for avoiding the constraints coming from the BBN, as well as the bounds connected with dark matter production. We have shown that the scalar field can start to evolve after the BBN leaving all the fundamental constants unchanged during the primordial production of the light elements. In this case, the coupling to matter is less suppressed which allows for the experimental searches for this kind of dark matter coupled to mass.

The authors are grateful to S. Sibiryakov, D. Gorbunov for valuable discussions and useful comments. This work was supported by Russian Science Foundation Grant $16$-$12$-$10494$.
\label{sec:Conclusions}

\section*{Appendix}
In this Appendix, we discuss briefly the Standart BBN model. More detailed review is given in \cite{BBN}.

The evolution of the fractional neutron abundance $X_n$ is described by  the balance equation,
\be
\frac{dX_n(t)}{dt}=\lambda_{pn}(t)\left(1-X_n(t)\right)-\lambda_{np}(t)X_n(t),
\ee

where $\lambda_{pn}$ is the summed rate of the reactions which convert neutrons to protons,
\be
\lambda_{pn}=\lambda(n\nu_e\rightarrow p e^-)+\lambda(n e^+\rightarrow p \bar{\nu_e})+\lambda(n\rightarrow p e^-\bar{\nu_e}),
\ee
and $\lambda_{np}$ is the rate of the reverse reactions which convert protons to neutrons is given by the detailed balance,
\be
\lambda_{pn}=\lambda_{np}\;e^{-\Delta m/T(t)}, \quad \Delta m \equiv m_n-m_p=1.293\:\text{MeV}. 
\ee 

The equilibrium solution is obtained by setting $dX_n(t)/dt=0$,
\be
X^{eq}_n(t)=\frac{\lambda_{pn}(t)}{\Lambda(t)}=\left(1+e^{\Delta m/T(t)}\right)^{-1}, \;\;\; \Lambda\equiv \lambda_{pn}+\lambda_{np},
\ee
while the general solution is
\be
\label{eq:gensol}
\left\{
\begin{array}{lcl}
{X_n(t)=\int\limits_{t_i}^t dt^\prime I(t, t^\prime)\lambda(t^\prime)+I(t, t_i)X_n(t_i)\;,} \\
{I(t, t^\prime)=\exp\left(-\int\limits_{t^\prime}^t dt^{\prime\prime}\Lambda(t^{\prime\prime})\right)\;.}
\end{array}
\right.
\ee
Since the rates $\lambda_{pn}$ and $\lambda_{np}$ are very large at early times, $I(t, t_i)$ will be negligible for a suitably early choice of the initial epoch, hence the initial value of the neutron abundance $X_n(t_i)$ plays no role and thus does not depend on any particular model of the very early Universe. For the reason, $t_i$ may be replaced by zero and the above expression simplifies to
\be
X_n(t)=\int\limits_0^t dt^\prime I(t, t^\prime)\lambda(t^\prime)=\frac{\lambda_{pn}(t)}{\Lambda(t)}-\int\limits_0^t I(t, t^\prime)\frac{d}{dt^\prime}\left(\frac{\lambda_{pn}(t^\prime)}{\Lambda(t^\prime)}\right). 
\ee

The neutron freeze-out temperature can be evaluated from dimensional considerations which give
\be
T_{fr}\sim \left(\frac{g_{\rho}^{1/2}}{G_F^2 M_P}\right)^{1/3}\sim 1 \: \text{MeV},
\ee
where $G_F$ is the Fermi constant.

The rates of reactions is given by the formula
\be
\lambda(n\nu_e\rightarrow p e^-)=\lambda(n e^+\rightarrow p\bar{\nu_e})=A\;T^3\left(24\;T^2+12\;T\Delta m+2\;(\Delta m)^2\right).
\ee
The constant A is related to $\tau_n$ as follows,
\be
\frac{1}{\tau_n}=0.0158\;A\;(\Delta m)^5.
\ee

Hence the total reaction rate can be expressed in terms of the neutron lifetime as,
\be
\lambda_{pn}(t)\simeq 2\lambda(n\nu_e\rightarrow p e^-)=\frac{a}{\tau_n y^5}(12+6y+y^2), \quad y\equiv \frac{\Delta m}{T},\quad a=253.
\ee

The integrating factor in (\ref{eq:gensol}) can now be calculated,
\be
I(y, y^\prime)=\exp\left(-\int^y_{y^\prime} dy^{\prime\prime}\frac{dt^{\prime\prime}}{dy^{\prime\prime}} \Lambda(y^{\prime\prime}) \right)=\exp(K(y)-K(y^\prime)),
\ee
where
\be 
K(y)=-b\left(\left(\frac{4}{y^3}+\frac{3}{y^2}+\frac{1}{y}\right)+\left(\frac{4}{y^3}+\frac{1}{y^2}\right)e^{-y}\right),
\ee
and
\be
b=a\left(\frac{45}{4\pi^3 g_\rho} \right)^{1/2}\frac{M_P}{\tau_n (\Delta m)^2}.
\ee

The neutron abundance is therefore,
\be
X_n(y)=X_n^{eq}(y)+\int\limits_0^y dy^\prime e^{y^\prime}(X_n^{eq}(y^\prime))^2 \exp(K(y)-K(y^\prime)).
\ee

By the time of the BBN the neutron abundance surviving at freeze-out has been depleted by $\beta$-decay to,
\be 
X_n(t_{BBN})\simeq X_n(y\rightarrow\infty)e^{-t_{BBN}/\tau_n}.
\ee
Nearly all of these surviving neutrons are captured in $^4He$ because of its large binding energy ($\Delta_{^4He}=28.3$ MeV). Hence the resulting mass fraction of helium is simply given by
\be
X_{^4He}\simeq 2X_n(t_{BBN}).
\ee
\label{sec:BBN}

\bibliography{MyBib}{}

\end{document}